\documentclass[preprint,5p,times,twocolumn]{elsarticle}

\usepackage{amsmath}
\usepackage{mathtools,amssymb}
\usepackage{color}
\usepackage{graphicx}
\usepackage{caption}
\usepackage{subcaption}
\usepackage{caption}
\usepackage{multicol}
\usepackage{bicaption}
\usepackage[hidelinks]{hyperref}
\usepackage{comment}

\usepackage[none]{hyphenat}

\journal{Extreme Mechanics Letters}
\biboptions{sort&compress}

\begin{document}

\begin{frontmatter}

\title{Experimental realization of tunable Poisson's ratio in deployable origami metamaterials}


\author[1]{Diego Misseroni\corref{cor}}
\author[2]{Phanisri P.\,Pratapa}
\author[3]{Ke Liu}
\author[4]{Glaucio H.\,Paulino\corref{cor2}}

\address[1]{Department of Civil, Environmental and Mechanical Engineering, University of Trento, Trento 38123, Italy}
\address[2]{Department of Civil Engineering, Indian Institute of Technology Madras, Chennai 600036, TN, India}
\address[3]{Department of Advanced Manufacturing and Robotics, Peking University, Beijing 100871, China}
\address[4]{Department of Civil and Environmental Engineering, Princeton University, Princeton, New Jersey, 08544, USA}
\cortext[cor]{Corresponding Author (\it diego.misseroni@unitn.it) }
\cortext[cor2]{Corresponding Author (\it gpaulino@princeton.edu) }


\begin{abstract}

Origami metamaterials are known to display highly tunable Poisson's ratio values depending on their folded state. Most studies on the Poisson effects in deployable origami tessellations are restricted to theory and simulation. Experimental realization of the desired Poisson effects in origami metamaterials requires special attention to the boundary conditions to enable deployable nonlinear deformations that give rise to tunability. In this work, we present a novel experimental setup suitable to study the Poisson effects in 2D origami tessellations that undergo simultaneous deformations in both the applied and transverse directions. The setup comprises a gripping mechanism, which we call a \textit{Saint-Venant fixture}, to eliminate Saint-Venant end effects during uniaxial testing experiment. 
Using this setup, we conduct Poisson's ratio measurements of the Morph origami pattern whose configuration space combines features of the Miura-ori and Eggbox parent patterns. We experimentally observe the Poisson's ratio sign switching capability of the Morph pattern, along with its ability to display either completely positive or completely negative values of Poisson's ratio by virtue of topological transformations. To demonstrate the versatility of the novel setup we also perform experiments on the standard Miura-ori and the standard Eggbox patterns. Our results demonstrate the agreement between the theory, the simulations, and the experiments on the Poisson's ratio measurement and its tunability in origami metamaterials. The proposed experimental technique can be adopted for investigating other  tunable  properties  of  origami  metamaterials in static and in dynamic regimes, such as finite-strain  Poisson’s  ratios,  elastic thermal  expansion, and wave propagation control.  
\end{abstract}


\end{frontmatter}


\section{Introduction}

Origami engineering attracts exponentially increasing interest among researchers due to its interdisciplinarity combining knowledge from mathematics, material science, computational and experimental mechanics. In fact, the optimal design of origami-inspired metamaterials requires the development of cutting-edge analytical modeling, advanced numerical simulations and sophisticated experimental tools~\citep{filipov2015origami,yasuda2019origami}. The extreme deployability of origami patterns, usually governed by just one kinematic parameter, makes them a source of inspiration for designing innovative structures with a wide range of applications in several fields, including physics, engineering, material science, space structures, chemistry, biology, architecture, robotics, biomimetic engineering~\citep{li2019architected, meloni2021engineering}. The static and dynamical mechanical properties of origami metamaterials mainly depend on the geometrical properties of their microstructure~\citep{christensen2015vibrant, bertoldi2017flexible}. One of the most investigated mechanical properties from a theoretical, numerical and experimental point of view is the Poisson's effect of artificially-made architected materials~\citep{clausen2015topology}. Since its  introduction~\citep{lakes1987foam,evans1991auxetic}, huge effort has been put by 
researchers to invent new metamaterials exhibiting an \textit{auxetic} behavior, usually by changing the geometrical design of their constituent 2D~\citep{bertoldi2010negative,cabras2014auxetic,li2019novel,bacigalupo2020chiral,auricchio2019novel,czajkowski2022conformal,coulais2016periodic,dos2021design} and 3D unit cell~\citep{babaee20133d,cabras2016class,baldi2022three}. Within this framework, origami metamaterials, being able to vary their geometrical configuration continuously from folded to unfolded states, and vice-versa, lead to an extreme tunability of their mechanical response depending on the folding state~\citep{schenk2013geometry, eidini2015unraveling, yasuda2015reentrant, boatti2017origami, pratapa2018bloch, liu2020origamiimperfection}. For instance, it has been theoretically and numerically demonstrated that the Poisson's ratio ($\nu$) of origami patterns can be always negative (Miura-ori~\citep{schenk2013geometry}, $\nu \in (-\infty,0]$), positive (standard Eggbox~\citep{schenk2013geometry}, $\nu \in [0,+\infty)$), or even switches from positive to negative and vice-versa (Morph pattern~\citep{pratapa2019geometric}, $\nu \in \mathbb{R}$). 
The standard Miura-ori is a developable origami pattern made from parallelogram shaped panels, and is a deployable single degree of freedom (SDOF) system when all the panels are considered rigid~\citep{schenk2013geometry}. The standard Eggbox is also made from the parallelogram shaped panels and is a non-developable SDOF origami pattern when the panels are rigid~\citep{schenk2013geometry}. The Morph origami pattern is a recently discovered metamaterial~\citep{pratapa2019geometric} which can undergo a change in the mountain-valley assignment of its creases, unlike the standard Miura-ori or standard Eggbox patterns where crease topology is fixed. A single Morph pattern can transform into hybrid states formed from a heterogeneous combination of unit cells in contrasting modes, and therefore, can exhibit a wide range of trends associated with tunable Poisson's ratios~\citep{pratapa2021reprogrammable}. Similar to standard Miura-ori and standard Eggbox, the Morph pattern is also an SDOF system when all the panels are assumed to be rigid. In practice, however, the panels have finite rigidity and therefore pose a problem in experimental realization of the origami metamaterial properties that are theoretically described using rigid panel assumption. 
The actual realization of origami-based devices required a deep understanding on the factors that can induce discrepancies from the mechanical properties predicted from the theory and those found experimentally, such as the effect of boundary conditions~\citep{jules2021curving}, manufacturing process~\citep{park2019review,ning2018assembly,chen2020kirigami}, panels rigidity and thickness accommodation~\citep{lang2018review,zhu2019efficient}. Although several studies have dealt with the experimental validation of Poisson's ratio effect of architected metamaterials and kirigami-based structures~\citep{naghavi2020fish,cabras2019micro,morvaridi2021hierarchical,fernandes2021mechanically,wang2021structured}, only few provide an equivalent investigation in the case of origami metamaterials~\citep{lin2020folding,he2020programming,wang2020modulation,yasuda2015reentrant}. Here, we propose a new experimental setup to determine the mechanical properties of deployable origami metamaterials during the entire folding/unfolding process. The gripping system, called \textit{Saint-Venant fixture}, has been carefully designed to allow the complete deployment of the tested metamaterials in the transverse direction, preventing sample frustration at the constrained regions.
Usually, uniaxial testing experiments on metamaterials require very long samples to ensure a sufficiently large area in the centre of the sample without boundary effects~\citep{morvaridi2021hierarchical}. In fact, standard gripping devices induce stress concentrations in the bonded areas, giving the tested metamaterial a dog bone shape that is symptom of a non-uniform transverse deformation. 
The novel setup proposed here breaks the limits of standard gripping mechanisms eliminating Saint-Venant end effects and thus permitting the testing of short samples. 
We have proved the efficiency of the new gripping mechanism by performing uniaxial testing experiments on several origami patterns that exhibit great Poisson's ratio variability. In particular, we have tested a Morph pattern in four different hybrid states, the standard Eggbox, and the standard Miura-ori patterns. Videos of the Poisson's ratio experiments performed on the Morph pattern (Movie SM1) and the Standard Miura-ori and Eggbox patterns (Movie SM2) are available as complementary material. Through our proposed experimental setup, we describe how the deployable (SDOF) deformations of the origami patterns can be realized despite the apparent non-rigidity of the panels that are fabricated. We then apply it to obtain the Poisson's ratios of the origami patterns at a continuum of folded (deformed) states to validate their theoretically predicted tunable nature.
This article is organized as follows. We first review the theoretical background to describe the Poisson's ratios and their tunability parameter of the origami patterns under consideration. We then present the experimental setups that will be used for the uniaxial tests required for the Poisson's ratio study. We also provide details on the fabrication of the origami specimens used for the study. Finally we present a comparison and discussion of the tunability of Poisson's ratios obtained from theory, experiment, and simulation performed by MERLIN software~\citep{liu2017nonlinear,liu2018merlin2}.


\section{Poisson's ratio of origami metamaterials}~\label{sec:theory}
Poisson's ratio for a material is defined as the negative of the ratio of transverse and applied strains, during a uniaxial deformation experiment. Since origami metamaterials can undergo large deformations due to nonlinear folding mechanisms, we calculate a tangential Poisson's ratio at each folded state by taking the ratio of the infinitesimal (Cauchy) strains along the two directions of the tessellation. Using the infinitesimal strain theory, we define axial strain of a segment with length $S$ as $\varepsilon_S = \mathrm{d}S/S$. Noting that the length $S$ is a function of the folded state represented by a folding-angle $\psi$, we can write $\varepsilon_S (\psi) = (\mathrm{d}S/\mathrm{d}\psi)(\mathrm{d}\psi/S)$. The dependency of the strains and correspondingly the Poisson's ratios of the origami metamaterials on the folding-angle $\psi$ gives rise to \emph{tunability}.

\begin{figure*}[ht]
	\centering
	\includegraphics[width=0.98\textwidth]{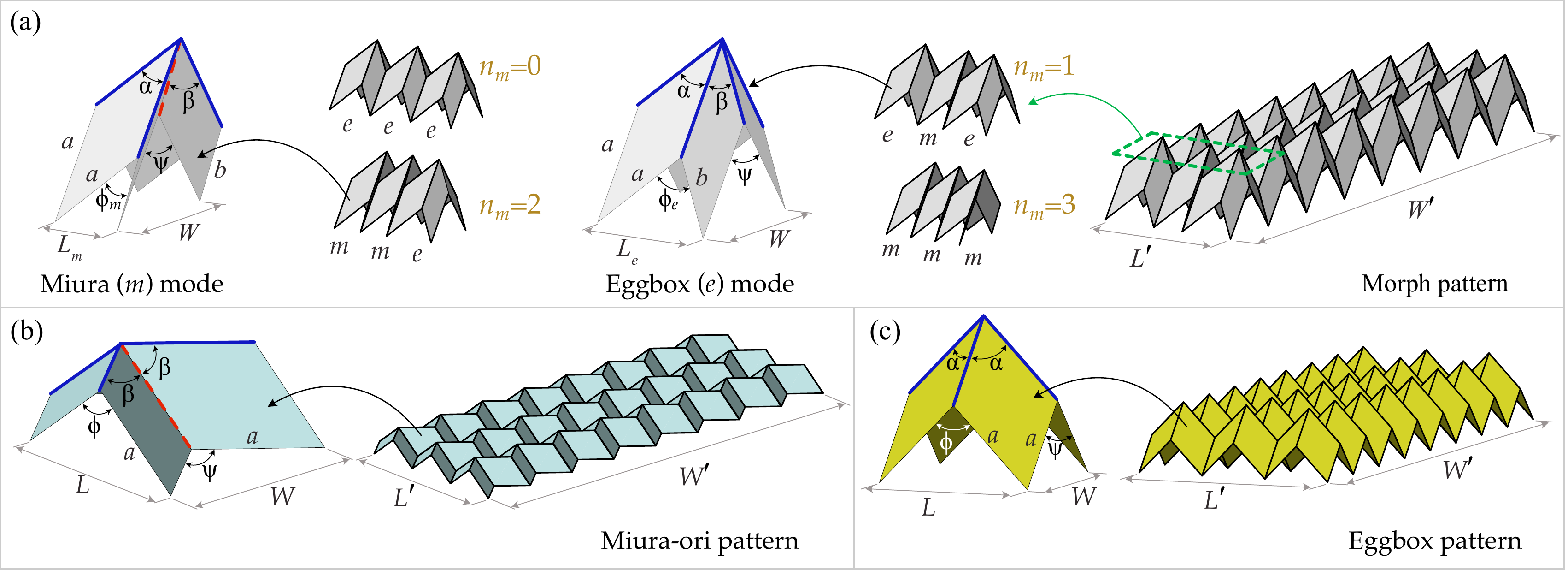}
	\caption{Geometry of (a) Morph origami patterns, (b) standard Miura-ori, and (c) standard Eggbox. The mountain and valley creases of the isolated unit cells are shown in solid blue and dashed red lines, respectively. }
	\label{Fig_Geometry}
\end{figure*}

In this section, we review the mathematical expressions for the Poisson's ratios of the Morph, the standard Miura-ori, and the standard Eggbox patterns. Schematics of these patterns and their unit cell geometries are shown in Fig.~\hyperref[Fig_Geometry]{\ref*{Fig_Geometry}}. We denote the dimensions of the two-dimensional patterns as $W'$ and $L'$ and the number of cells along the corresponding dimensions as $n_w$ and $n_\ell$, respectively. 

\subsection{Morph pattern}
Let $\alpha$ and $\beta$ denote the panel angles of the Morph unit cell. The panel dimensions, denoted as $a$ and $b$, are constrained to satisfy $b=\zeta a$ in order to ensure the orthorhombic nature of the unit cell, where $\zeta=\cos\alpha/\cos\beta$. Unlike standard Miura-ori or standard Eggbox patterns, the Poisson's ratio of the Morph pattern could be different from that of its unit cell depending on the hybrid state it exists in. Hybrid states are heterogeneous configurations of the Morph pattern that consist of the Miura mode and the Eggbox mode cells simultaneously. Figure~\hyperref[Fig_Geometry]{\ref*{Fig_Geometry}(a)} shows a Morph pattern in its hybrid state consisting of two Eggbox mode cells and one Miura mode cell. Other possible hybrid states can also be obtained by changing the proportion of the number of Miura mode cells ($n_m$) or the number of Eggbox mode cells ($n_\ell - n_m$) along the $L'$ dimension of the structure (see Fig.~\hyperref[Fig_Geometry]{\ref*{Fig_Geometry}(a)}). 

For $\psi \in \left[0, 2\beta\right]$, the dimensions of the pattern are calculated as 
\begin{align}
W'(\psi)&=n_w W(\psi) = 2n_w a \sin(\psi/2) \label{Eq:Wp}\,, \\
L'(\psi)&=(n_\ell - n_m) L_e + n_m L_m \nonumber \\
    &=(n_\ell - n_m)a \Phi_e(\psi)+n_m a \Phi_m(\psi) \,, \label{Eq:Lp}
\end{align}
where
\begin{align}
\Phi_e(\psi) &= \sqrt{1+\zeta^2-2\zeta\cos\phi_e} \,, \label{Eq:PHIe}\\
\Phi_m(\psi) &= \sqrt{1+\zeta^2-2\zeta\cos\phi_m} \,, \label{Eq:PHIm}
\end{align}
with $\phi_e=\phi_e(\psi)$ and $\phi_m=\phi_m(\psi)$ given by
\begin{align}
\phi_e(\psi)&=\cos^{-1}\bigg(\frac{\cos\alpha}{\cos(\psi/2)}\bigg)+\cos^{-1}\bigg(\frac{\cos\beta}{\cos(\psi/2)}\bigg) \,, \label{Eq:phi_e} \\
\phi_m(\psi)&=\cos^{-1}\bigg(\frac{\cos\alpha}{\cos(\psi/2)}\bigg)-\cos^{-1}\bigg(\frac{\cos\beta}{\cos(\psi/2)}\bigg) \,.\label{Eq:phi_m}
\end{align}
The Poisson's ratio of the Morph pattern in a hybrid state, is given by
\begin{align}
\nu_{WL}^H(\psi)&=-\frac{\varepsilon_{L'}(\psi)}{\varepsilon_{W'}(\psi)} = p(\psi) q(\psi)\,, \label{Eq:nuGlobal0}
\end{align}
where
\begin{align}
p(\psi)=\frac{\cos\alpha\sin^2\frac{\psi}{2}}{\left[(n_\ell - n_m)\Phi_e(\psi)+n_m\Phi_m(\psi)\right]\cos\beta\cos^3\frac{\psi}{2}} \,, \label{Eq:nuGlobal1}
\end{align}
and
\begin{align}
q(\psi)=\bigg[& \frac{\cos\alpha}{\sin((\phi_e+\phi_m)/2)}\bigg(\frac{(n_\ell - n_m)\sin\phi_e}{\Phi_e(\psi)}+\frac{n_m\sin\phi_m}{\Phi_m(\psi)}\bigg) \nonumber
\\&+ \frac{\cos\beta}{\sin((\phi_e-\phi_m)/2)}\bigg(\frac{(n_\ell - n_m)\sin\phi_e}{\Phi_e(\psi)}-\frac{n_m\sin\phi_m}{\Phi_m(\psi)}\bigg) \bigg] \,. \label{Eq:nuGlobal2}
\end{align}
Depending on the choice of $n_\ell$ and $n_m$, the Morph pattern can exhibit any real value of Poisson's ratio, i.e.\,$-\infty<\nu_{WL}^H(\psi)<\infty$. Note that Eqn.~\ref{Eq:nuGlobal0} also gives the Poisson's ratio of the homogeneous states of the Morph pattern when ($n_\ell - n_m$) or $n_m$ is equal to $0$.

\subsection{Standard Miura-ori pattern}
Let $\beta$ denote the panel angle of the Miura-ori unit cell. For $\psi \in \left[0, 2\beta\right]$, the dimensions of the Miura-ori pattern are calculated as 
\begin{align}
W'(\psi)&=n_w W(\psi) = 2n_w a \sin(\psi/2) \label{Eq:W}\,, \\
L'(\psi)&=n_\ell L(\psi) = 2n_\ell a \sin(\phi/2) \,, \label{Eq:L}
\end{align}
where
\begin{align}
\phi = \phi(\psi)=  2\sin^{-1}\bigg(\frac{\cos\beta}{\cos(\psi/2)}\bigg)\,. \label{Eq:phi_Miura}
\end{align}
The Poisson's ratio of a Miura-ori unit cell (and also the pattern) is given by
\begin{align}
\nu_{WL}^M(\psi)&=-\frac{\varepsilon_{L}(\psi)}{\varepsilon_{W}(\psi)}=-\tan^2(\psi/2)\,. \label{Eq:nu_Miura}
\end{align}
Note that $-\infty <\nu_{WL}^M(\psi)\leq 0$.

\subsection{Standard Eggbox pattern}
Let $\alpha$ denote the panel angle of the Eggbox unit cell. For $\psi \in \left[0, 2\alpha\right]$, the dimensions of the Eggbox pattern are calculated using the same equations (Eqns.~\ref{Eq:W} and~\ref{Eq:L}) as that of Miura-ori except for the expression of angle $\phi$, which is given by
\begin{align}
\phi = \phi(\psi)=  2\cos^{-1}\bigg(\frac{\cos\alpha}{\cos(\psi/2)}\bigg)\,. \label{Eq:phi_Eggbox}
\end{align}
The Poisson's ratio of an Eggbox unit cell (and also the pattern) is given by
\begin{align}
\nu_{WL}^E(\psi)&=-\frac{\varepsilon_{L}(\psi)}{\varepsilon_{W}(\psi)}=\frac{\cos^2\alpha\tan^2(\psi/2)}{\cos^2(\psi/2)-\cos^2\alpha}\,. \label{Eq:nu_Eggbox}
\end{align}
Note that since $\psi\leq 2\alpha$, the denominator in the above expression is always non-negative, and therefore, \mbox{$0\leq \nu_{WL}^E(\psi) <\infty$}.


\section{Fabrication of the origami metamaterials}

\begin{figure*}[!ht]
	\centering
	\includegraphics[width=0.99\linewidth]{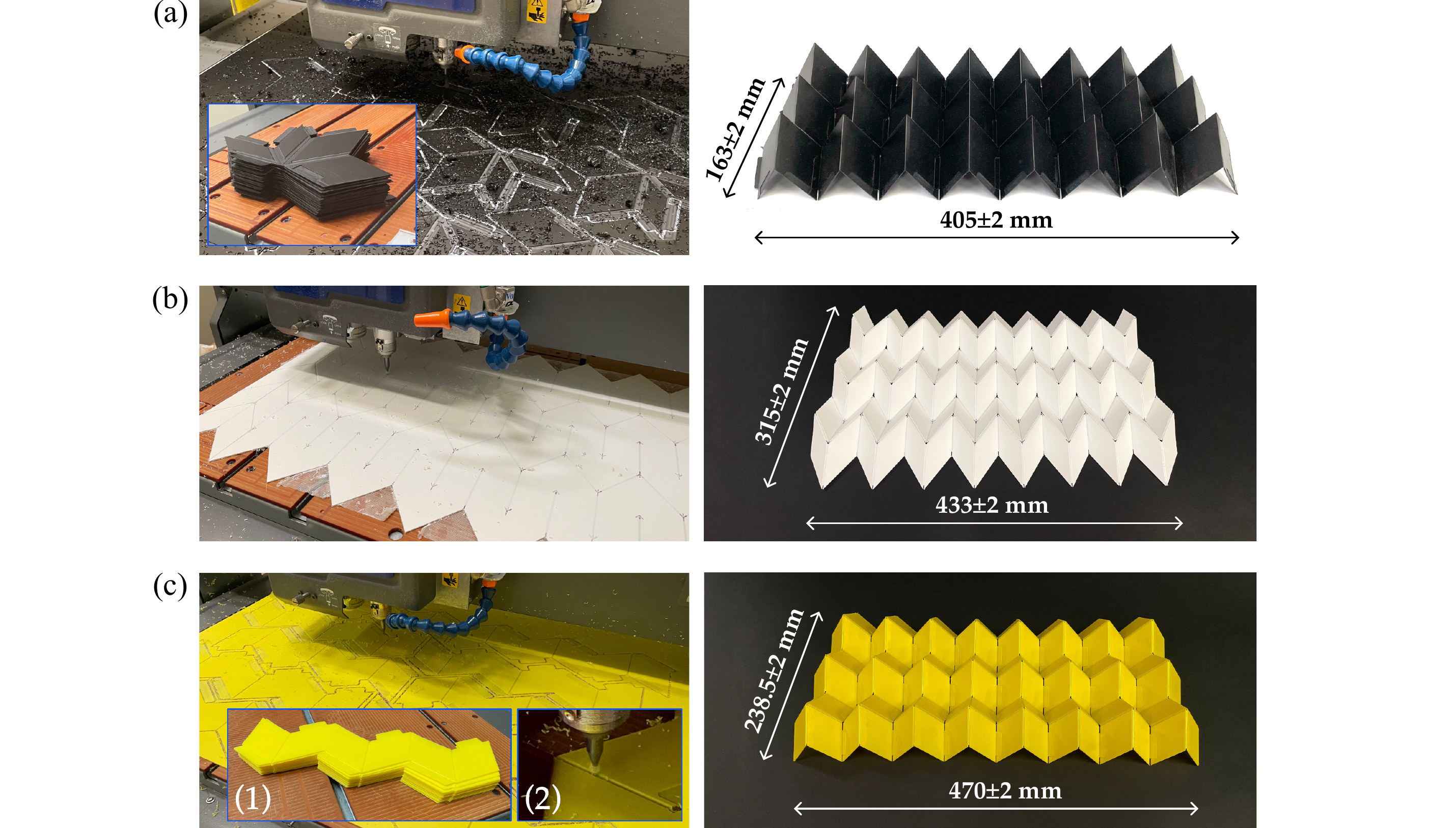}
	\caption{Milling machine in action during the manufacturing the origami metamaterials. (a) Milling of the Morph pattern unit cells depicted in the left inset.  Right part: the Morph pattern after its assemblage. (b) The standard Miura-ori pattern, shown in the right part, is obtained by folding a 1 mm thick Polypropylene sheet milled with the CNC machine. The dimension of the sample in its flat configuration is 514$\times$538 mm. (c) Realization of the strips 3$\times$1/2 unit cells used to assemble the standard Eggbox pattern. The inset (1) shows a stack of 3$\times$1/2 strips prior the  assemblage, while the inset (2) depicts the realization of the folding line by a ball end mill (radius 1 mm) mounted on the milling machine.  Right part: Eggbox pattern after its assemblage.}
	\label{Fig:Manufacturing}
\end{figure*}

We have manufactured three types of origami metamaterials made up of of  8$\times$3 unit cells, namely the Morph, the standard Miura-ori, and the standard Eggbox patterns, as shown in Figs.~\hyperref[Fig:Manufacturing]{\ref*{Fig:Manufacturing}(a)},~\hyperref[Fig:Manufacturing]{\ref*{Fig:Manufacturing}(b)} and~~\hyperref[Fig:Manufacturing]{\ref*{Fig:Manufacturing}(c)}, respectively. 

The Morph tessellation was created by assembling single unit cells obtained by cutting a black 1 mm thick Polypropylene sheet with a 3-axes CNC milling machine (EGX-600 by Roland, accuracy 10 $\mu$m), as shown in Fig.~\ref{Fig:Manufacturing}(a). The unit cell is generated by folding its flat configuration and using just one bond. Specific seats and extensions were realized on the unit cell panels to permit a rapid assemblage of the whole tessellation shown in the right part of Fig.~\ref{Fig:Manufacturing}(a). 
The standard Miura-ori pattern was generated from a one-piece of white Polypropylene sheet (1 mm thick) by the same milling machine used to create the Morph pattern, as shown in Fig.~\ref{Fig:Manufacturing}(b). Mountain and valley folding lines were created respectively on the top and bottom surface of the Polypropylene sheet to limit conflict issues during the folding process due to panels thickness. The dimension of the sample in its flat configuration is 514$\times$538 mm.  The standard Eggbox pattern was obtained by assembling the strips of  3$\times$1/2 unit cells depicted in the inset (1) of Fig.~\ref{Fig:Manufacturing}(c).
Each strip was milled from a yellow 1 mm thick Polypropylene sheet by the milling machine. Specific seats and extensions were designed on the panels of the strips to permit the perfect assembly of the 2D tessellation shown in the right part of Fig.~\ref{Fig:Manufacturing}(c) and to make the arrangement highly modular. 
The folding lines of all the investigated origami patterns were realized by a ball end mill (radius 1 mm) mounted on the milling machine and had an average thickness of \mbox{0.25 $\pm$ 0.05 mm} (e.g. see inset (2) of Fig.~\ref{Fig:Manufacturing}(c)). 
The bonded junctions between each element of the tessellation were obtained using the Loctite$^{\footnotesize\textregistered}$ 406  (by Henkel) in combination with the Loctite$^{\footnotesize\textregistered}$ 770 primer (by Henkel).
With reference to the symbols reported in Fig.~\ref{Fig_Geometry}, the geometrical parameters of the investigated origami patterns are  $\alpha=60^\circ$, $\beta=40^\circ$, $n_w=8$, and $n_\ell=3$. The dimensions of the panels sides are  \mbox{$a=49.2$ mm}, \mbox{$b=32.1$} mm for the Morph pattern, and $a=50$ mm for the standard Miura-ori and the standard Eggbox patterns. The dimensions of each assembled pattern at the rest configuration are reported in Fig.~\ref{Fig:Manufacturing} (right part). We note that the Poisson's ratios of the origami metamaterials discussed in this work are independent of the actual length dimensions ($a$ or $b$) of the panels.


\section{Experimental setup for the Poisson's ratio study}\label{setup}
\begin{figure*}[ht!]
	\centering
	\includegraphics[width=0.98\textwidth]{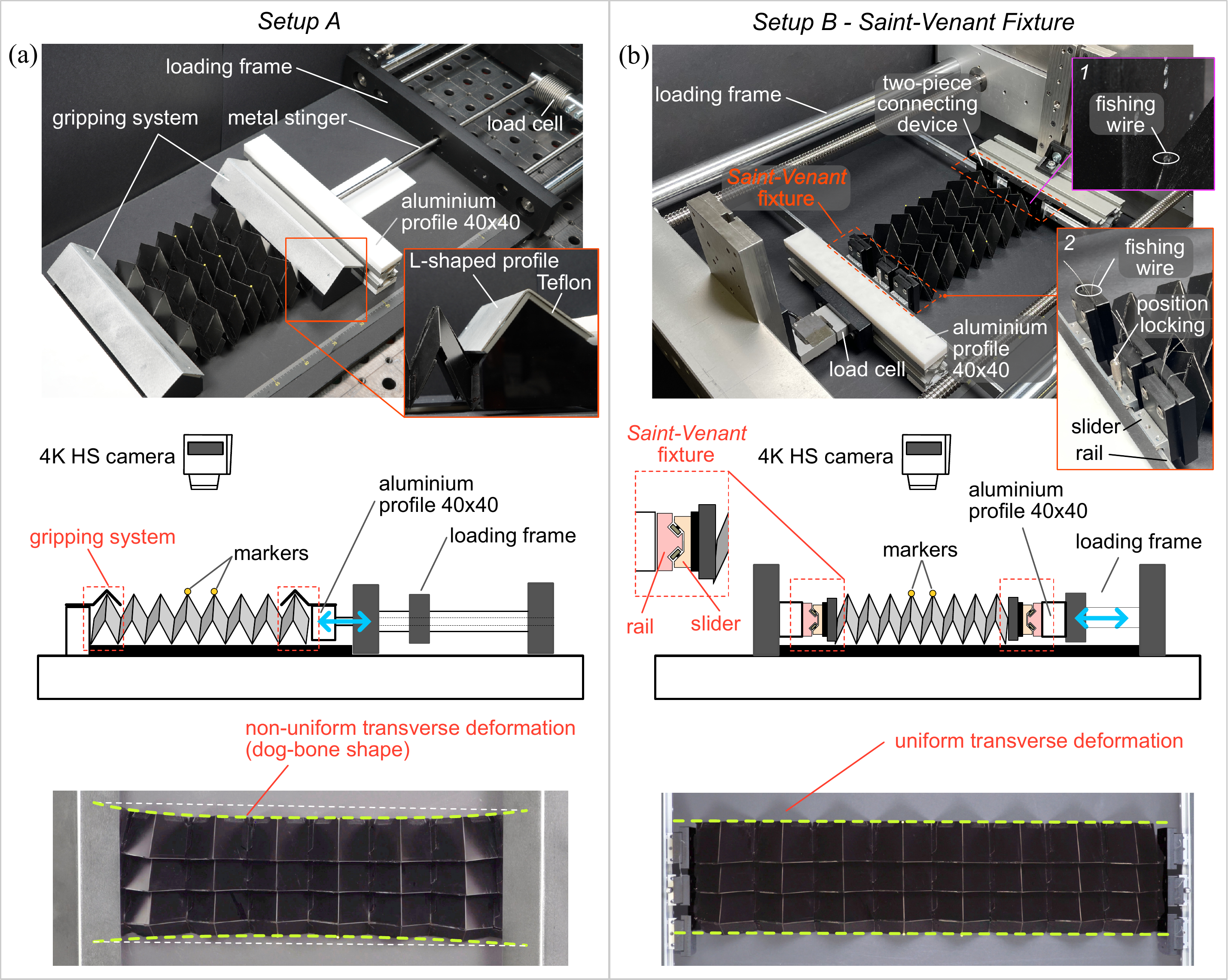}
	\caption{Details of the experimental setups designed and realized to perform the Poisson's ratio experiments. (a) Photo (upper part) and schematics (middle part) of \textit{setup A}. The sample is constrained with an L-shaped profile placed above the degree-4 vertex of the 3  unit cells of the sample end rows. Such constraints induce a non-uniform transverse deformation, as evidenced by the dog-bone shape assumed by the sample (bottom part). (b) Photo (upper part) and schematics (middle part) of \textit{setup B}. The sample is constrained with a linear slide system that comprises several sliders inserted into a rail, namely the \textit{Saint-Venant fixture}. The inset \textit{1} of the top figure depicts the connection between the panels of the pattern to the slider through the fishing wire, while the inset \textit{2} shows the detailed view of the linear slide system composing the \textit{Saint-Venant fixture}. The \textit{Saint-Venant fixture} allows a completely free sample deployment, avoiding the dog-bone shape (bottom part).  
	}
	\label{Fig:EXP:Setups}
\end{figure*}

We have performed systematic quantitative Poisson's ratio experiments on the Morph, the standard Miura-ori, and the standard Eggbox origami metamaterials to investigate the influence of boundary conditions on experimental measurements. To this purpose,  two types of setups, shown in Figs.~\hyperref[Fig:EXP:Setups]{\ref*{Fig:EXP:Setups}(a)} and~\hyperref[Fig:EXP:Setups]{\ref*{Fig:EXP:Setups}(b)} and hereafter called, respectively, \textit{setup A} and \textit{setup B},  were precisely designed to perform tensile tests on the deployable origami metamaterials. The setups were conceived to monitor the in-plane longitudinal and transverse displacements of the tested metamaterials. The whole experimental apparatus was arranged horizontally to prevent the gravitational effect and reduce out-of-plane instabilities that could arise during the execution of the tests. 
The main difference between the two setups concerns the fixture mechanisms (gripping system) used to connect the origami metamaterial to the loading frame machine that is thoroughly discussed in the following sections. The analysis of the experimental data, discussed in Section~\ref{results:discussion}, has shown that the choice of the gripping system influences Poisson's ratio measurements since it modifies the boundary conditions imposed on the origami pattern ends.

\subsection{Setup A}
This is the first setup we have developed. This setup was used to perform the first experimental campaign on the Morph origami metamaterial. The tested sample is constrained on the two shorter sides with a L-shaped aluminium profile located above the degree-4 vertex of the 3  unit cells of the pattern end rows, as shown in  Fig.~\ref{Fig:EXP:Setups}(a).  This fixing device was designed to prevent its relative longitudinal displacements with the constrained sample leaving  the transverse motion sufficiently free.  The L-shaped profiles used to constrain the origami patterns were covered with a 3 mm thick Teflon (ETFE) plate to reduce the friction with sample and were mounted, in turn, on two aluminium profiles 40x40 mm. The latter were used to fix one end of the metamaterial to an optical table (from Thorlabs) and to connect the other to a loading frame machine by a circular metal stinger of length 650 mm. A Teflon plate was also placed beneath the sample to reduce the friction with the PMMA plate that supports the pattern during the execution of the tests. The tensile tests were performed by applying a constant speed of 4 mm/s at one end of the metamaterial with a  Messphysik $\mu$-strain loading frame machine (0.01 $\mu$m stroke measurement resolution). Load $P$ and global longitudinal displacement $\delta$ were acquired with a  AEP TYPE F1-1kN load cell and with a displacement transducer  mounted internally to the testing machine. Such data were recorded with a \mbox{Ni-cDaq 9188} acquisition system interfaced with a PC through a code developed in LabVIEW 2020. 

\subsection{Setup B - Saint-Venant Fixture}
We have thoroughly revised and improved the gripping system of \textit{setup A} to mimic as much as possible the boundary conditions underlying the theoretical formulation and achieve best agreement between experiments and theoretically predicted Poisson's ratio. Photo and schematics of \textit{setup B}  are reported in Fig.~\ref{Fig:EXP:Setups}(b). The substantial difference with \textit{setup A}, concerns the replacement of the L-shaped aluminium profiles with a more sophisticated gripping mechanism that eliminates the sample's frustration at its constrained ends permitting the complete deployment of the origami metamaterial during the folding/unfolding process.  In this case, the gripping mechanism used to connect  the metamaterial ends to the two aluminium profiles 40x40 mm, one fixed on the optical table and one connected to the loading frame, consists of a linear slide system (purchased from MiSUMi Europe) composed of  several sliders inserted into a rail.   Each slider can slide without friction against a perfectly fitting rail constraint through four rolling bearings and is equipped with a position locking system. The locking system is used to block one of the sliders inserted in the rail to avoid rigid motion of the metamaterial during the execution of the uniaxial testing experiments.  This gripping system is very versatile since the number of the sliders can vary depending on the type of experiment and the dimension of the sample to be tested. The slide rail system permits the perfect fixing of the sample while leaving free the transverse movement during the folding/unfolding process. A two-pieces  (PMMA black) connecting device has been designed and realized with a CNC milling machine to allow an easy connection of the sample to each slider by a 0.65 mm in diameter fishing wire, as shown in the detail of Fig.~\ref{Fig:EXP:Setups}(b). The fishing wire is carefully tensed to keep the panels of the tested patterns connected to the sliders during the whole folding/unfolding process, as depicted in the inset \textit{1} of Fig.~\ref{Fig:EXP:Setups}(b). Such a device also permits a fine-tuning of the connection point of the sample to the slider in the vertical direction. The fine adjustment of the connection point is fundamental when: i)  the samples to be tested have different heights or ii) the experiments involve origami metamaterials that exhibit a change of height during the folding/unfolding process. The efficiency of the novel fixture is evident comparing the two snapshots reported in the bottom part of Fig.~\ref{Fig:EXP:Setups}. The same Morph pattern ($n_m$=2) is subjected to uniaxial testing experiments by using \textit{setup A} (left) and \textit{setup B} (right). The gripping system of \textit{setup A} induces a non-uniform transverse deformation to the pattern as indicated by its dog-bone shape.  By contrast, the novel fixture, called \textit{Saint-Venant fixture}, permits a free deployment of the tested origami pattern characterized by a uniform transverse deformation, avoiding stress concentration in the constrained region of the metamaterial and thus eliminating Saint-Venant end effects. 
In the case of \textit{Setup A} (\textit{standard} gripping system), the total specimen length must be at least four times longer than the length of the central area with uniform deformation to avoid the local effects of constraints (bottom part of Fig~\ref{Fig:EXP:Setups}a), which might be quite impractical from an experimental point of view. Conversely, the Saint-Venant fixture permits the testing of very short samples, which are reliable in the sense of representing a true periodic system without violating the underlying theoretical hypothesis (bottom part of Fig~\ref{Fig:EXP:Setups}b).

The tensile experiments on the origami metamaterials were carried out at a constant speed of 4 mm/s applied at one end of the sample by a  Messphysik MIDI-10 loading frame machine (0.05 $\mu$m stroke measurement resolution). During the execution of the experiments, load $P$ and global longitudinal displacement $\delta$ were acquired respectively with a DBSSM-100kg (by Leane International) load cell and with a displacement transducer mounted internally to the testing machine. Such data were recorded with a Ni-cDaq 9188 acquisition system interfaced with a PC through a code developed in LabVIEW 2020. By using the \textit{setup B}, we have performed the second experimental campaign on the Morph origami metamaterial. The obtained results have been compared with those attained with \textit{setup A}. Furthermore, we have exploited the new gripping mechanism to test the standard Eggbox and the standard Miura-ori origami metamaterials. The latter experiment would have been impossible by using \textit{setup A} since the standard Miura-ori origami can shift from a 3D to a 2D configuration and vice-versa. Therefore, during the unfolding process, namely moving towards a flat configuration, the Miura-ori pattern would slip out from the L-shaped profiles adopted in \textit{setup A} to constrain the sample.

\subsection{Digital image correlation and tracking method}
\begin{figure}[t] 
	\centering
	\includegraphics[width=1\linewidth]{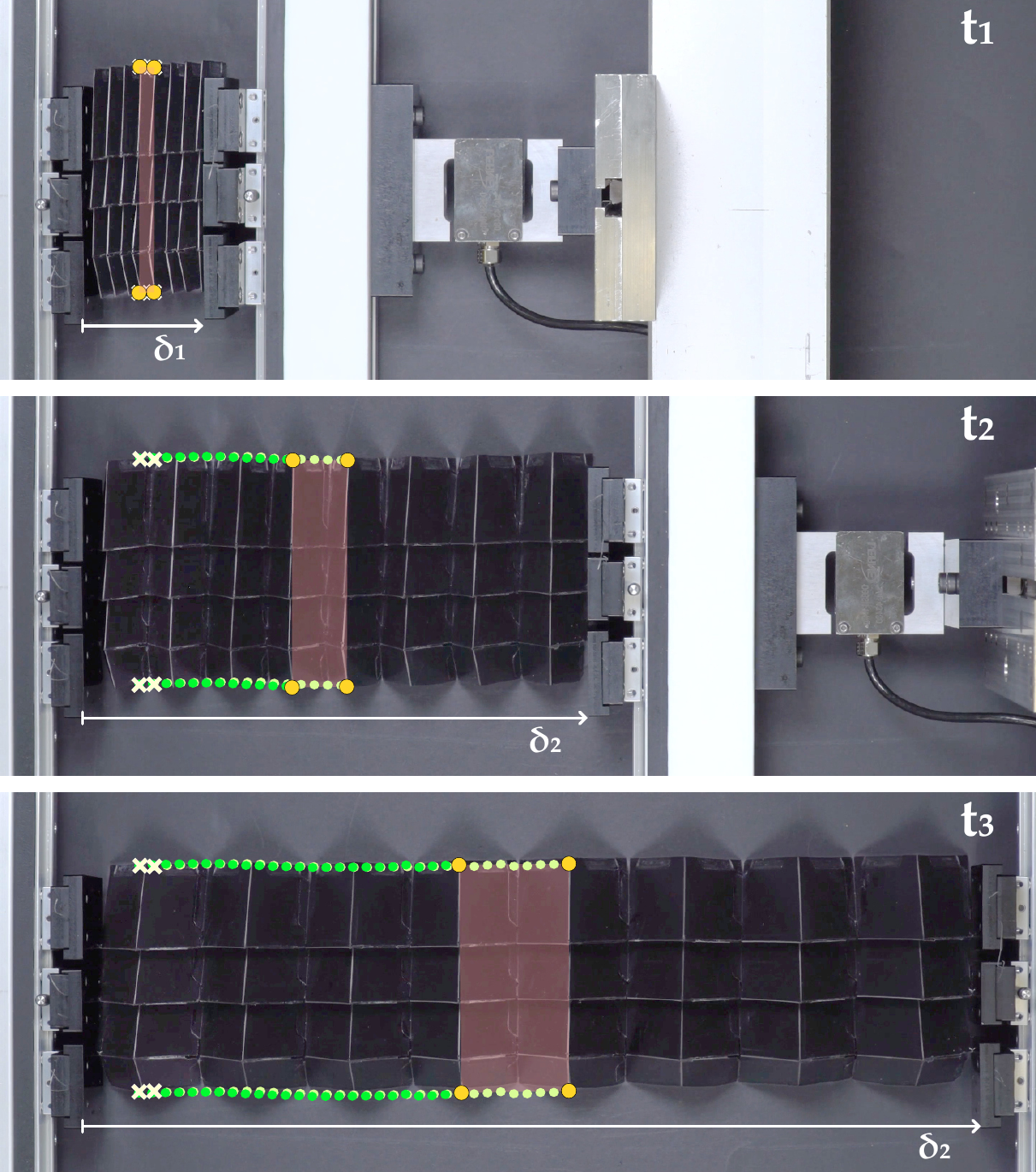}
	\caption{Snapshots documenting the deployable  mechanism of the Morph pattern ($n_m$=2) captured during the tensile tests at three different instants of time, $t_1$, $t_2$ and $t_3$. The trajectories of each marker obtained by the \textit{digital correlation and tracking analysis} are superimposed to each snapshot.}
	\label{Fig:Poisson:Tracking}
\end{figure}

We have placed a 4K HS Camera (Sony PXW-FS5) equipped with a G Master FE 100-400 mm orthogonal to the testing platform to record the experiments and monitor the motion of the spherical markers (1.5 mm in diameter) during the uniaxial tests.  We have mounted a telephoto lens to reduce distortion and create contrast between the foreground and background. The camera was synchronized with the Ni-cDaq data acquisition system. The markers define the rectangular regions in the middle of each pattern as highlighted in Fig.~\ref{Fig:Poisson:Tracking}. The location of the markers was carefully chosen to permit the correct Poisson's ratio measurement to avoid the local effects of constraints and ensure their visibility during the whole folding/unfolding process. The color of the markers was selected to enhance colour contrast with that of origami pattern to be tested.  Thus, we have used yellow markers on the black Morph pattern, red markers on the white standard Miura-ori pattern, and black markers on the yellow standard Eggbox pattern. Such choices facilitate the motion capture of the markers over time by a \textit{digital image correlation and tracking method}. The technique involves several steps. First, the experimental movies were post-processed in \textit{Final Cut Pro X} to enhance the visibility of the tracking markers by transforming the coloured frames to their green (background) and black (markers) versions. After this process, each snapshot appears green everywhere except for the black markers. 
 Then, we have developed a specific \textit{in-house} software in \textit{Mathematica 12.2} to perform the frame-by-frame post-processing analysis of the experimental movies (3840$\times$2160 pixels, 29.97 fps) for estimating the coordinates of the centre of mass of each marker over time. For the purpose of illustration, in Fig.~\ref{Fig:Poisson:Tracking} we show the trajectories of each marker as obtained by the \textit{digital correlation and tracking analysis} during the execution of a tensile test on the Morph pattern ($n_m$=2). 
 The marker trajectories are superimposed to the snapshots of the experiment. 
 Since the experiments are quasi-static, only 0.75 fps were considered to estimate the Poisson's ratio. 
Alongside the execution of the experiments, photos were taken with a Sony Alpha 9 camera and a Sony Alpha 6300 equipped with a Vario-Tessar T* FE 24-70 mm and Vario-Tessar T* E 16-70 mm lenses, respectively.


\section{Results and discussion}\label{results:discussion}

\begin{figure*}[htp!]
	\centering
	\includegraphics[width=0.98\textwidth]{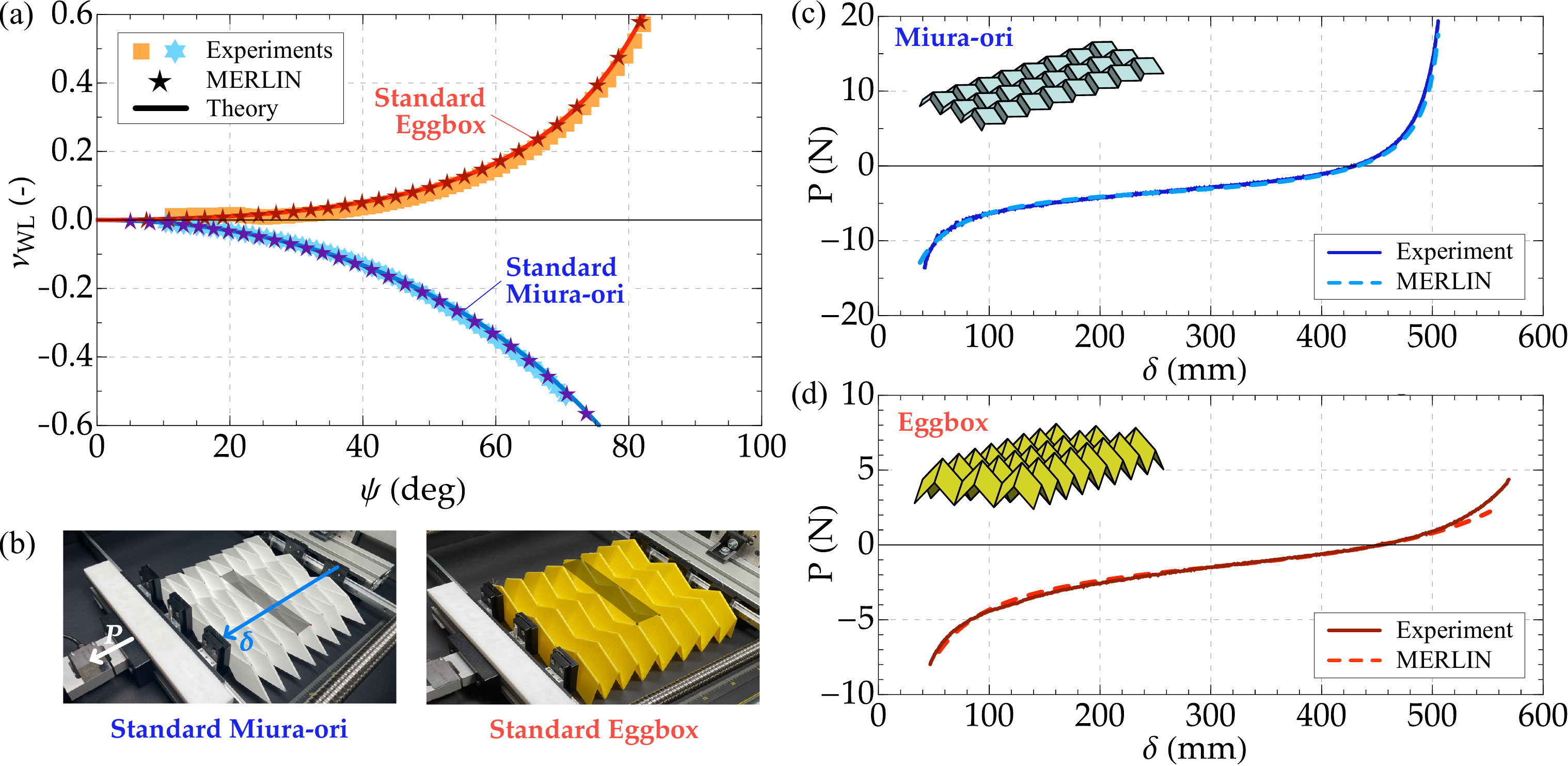}
	\caption{(a) Poisson's ratio experiments on the standard Miura-ori and the standard Eggbox patterns. In the graph, the experiments (markers) are compared with the theory (continuous lines), and the numerical simulations performed by MERLIN software (five-pointed star markers).  The experiments have been conducted by using the \textit{Saint-Venant fixture} (\textit{setup B}). (b) Photos of the investigated folded states. The highlighted areas in the middle of the patterns indicate the tracking regions used to evaluate the Poisson's ratio during the experiments. (c-d) Load/displacement curves recorded during the execution of the tests. The displacement $\delta$ represents the origami pattern length during the unfolding process. The experimental results (continuous lines) are superimposed to the numerical results obtained by MERLIN (dashed lines).}
	\label{Fig:Poisson:Stand}
\end{figure*}

We have tested all the fabricated origami patterns under tension to determine the evolution of their Poisson's ratio during an unfolding process. Noting the displacements of the markers (by the \textit{digital image correlation and tracking method}), we have evaluated the experimental Poisson's ratio through the forward difference scheme. Thus, the Poisson's ratio $\nu_i$, associated with the generic frame $i$ of the experimental movie, was computed via the relation
\begin{equation}
	\label{nu:i:exp}
	\nu_i=-\dfrac{W_i}{L_i}\dfrac{L_{i+1}-L_i}{W_{i+1}-W_i},
\end{equation}
where $L$ and $W$ are the width and the length of the unit cell, respectively (see Fig.~\ref{Fig_Geometry}). 
\begin{figure*}[ht!]
	\centering
	\includegraphics[width=0.98\textwidth]{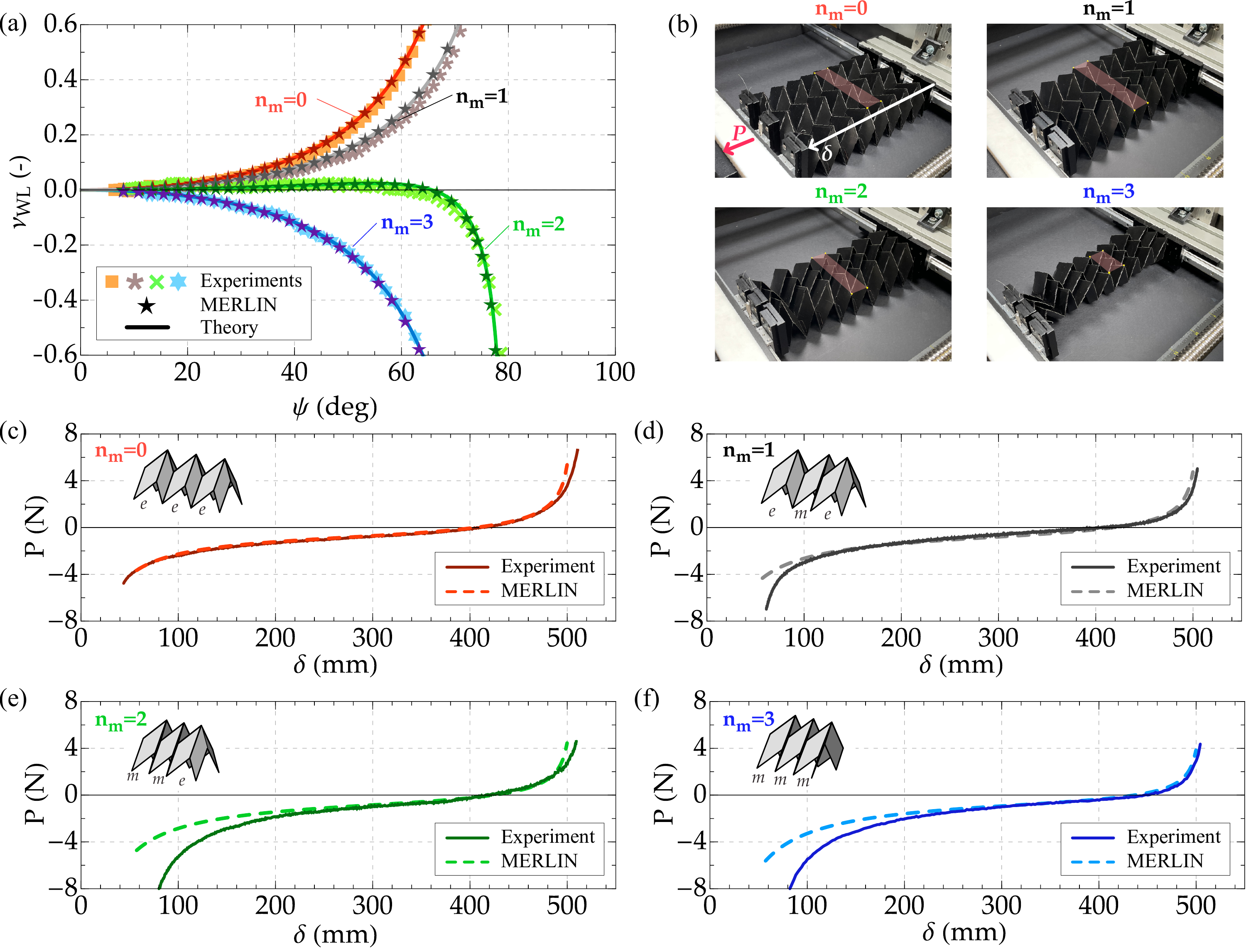}
	\caption{ (a) Poisson's ratio experiments on the four hybrid states of the Morph pattern ($n_m$=0,1,2,3) by using the \textit{Saint-Venant fixture}, namely the linear slide rail mechanism (\textit{setup B}). In the graph, the experiments (markers) are compared with the theory (continuous lines), and the numerical simulations performed with MERLIN software (five-pointed star markers). (b) The highlighted areas in the middle of the patterns indicate the tracking regions used to evaluate the Poisson's ratio during the experiments. (c-f) Load/displacement curves recorded during the execution of the tests. The displacement $\delta$ represents the origami pattern length during the unfolding process. The experimental results (continuous lines) are superimposed to the numerical results obtained by MERLIN (dashed lines). }
	\label{Fig:Poisson:Morph}
\end{figure*}
A five-points moving average was applied to slightly smooth the data obtained from Eqn.~\ref{nu:i:exp} and to produce the plots of Figs.~\ref{Fig:Poisson:Morph}(a),~\ref{Fig:Poisson:Stand}(a) and~\ref{Fig:Poisson:Morph:A}(a).   
This expedient was needed to purify the experimental data from spurious effects that inevitably arise during the execution of the tests.  Such effects are mainly due to friction and stick and slip phenomena between the bottom side of the origami sample and the experimental platform. 
We have tested the Morph pattern in 4 different configurations ($n_m=\{0,1,2,3\}$) to verify the tunability of the Poisson's ratio depending on the folded state by using both \textit{setup A} (first experimental campaign) and \textit{setup B} (second experimental campaign). The careful analysis of the experimental results  (Fig.~\ref{Fig:Poisson:Morph:A}(a)) and the movies of the experiments obtained by using \textit{setup A} indicates the  L-shaped gripping system as a main source of the discrepancy with the theoretically predicted Poisson's ratio. Such a constraint yields a dog-bone shape to the sample indicating a non-uniform transverse deformation during the folding/unfolding processes, as shown in Fig.~\ref{Fig:EXP:Setups}(a), bottom part. This spurious effect increases while the samples undergo large unfolding, namely for angles $\psi$ greater than 45$^\circ$. To circumvent this problem, we have designed and realized the \textit{setup B} that we used to perform the second experimental campaign on the Morph pattern and the experiments on the standard Miura-ori and Eggbox. We start with discussing the tests on the Morph pattern. 
The Poisson's ratio evolution as a function of the angle $\psi$ are reported in Fig.~\ref{Fig:Poisson:Morph}(a) for all the investigated hybrid states depicted in Fig.~\ref{Fig:Poisson:Morph}(b). The highlighted areas in the middle of the sample represent the tracking region adopted to evaluate the Poisson's ratio during the unfolding process. 
Note that $n_m=0$ and $n_m=1$ patterns exhibit a positive Poisson's ratio along the whole folding process. For $n_m=2$, the hybrid Morph pattern undergoes a change in the sign of the Poisson's ratio from positive to negative as it folds. For $n_m=n_\ell=3$, the Morph pattern is auxetic along the whole folding process. In Fig.~\ref{Fig:Poisson:Morph}(a), the experimental results (markers) are compared with the theoretical prediction (continuous lines) provided by Eqn.~\ref{Eq:nuGlobal0}, and the numerical simulations (see~\ref{app:MERLIN}) carried out by MERLIN software (five-pointed star markers). The colour of lines and markers indicate the investigated hybrid state, $n_m$=0 red, $n_m$=1 dark gray, $n_m$=2 green, and $n_m$=3 blue.
The analysis of the experiments shows an almost perfect coincidence between experiments, simulations and theory for all the investigated folded states and over the whole folding/unfolding processes (entire range of $\psi$). These results confirm the importance of the imposed experimental boundary conditions on Poisson's ratio measurements, especially when large deployment is achieved. A video recording of the experiments is also provided as supporting material (Movie SM1).
The graphs of Figs.~\hyperref[Fig:Poisson:Morph]{\ref*{Fig:Poisson:Morph}(c)},~\hyperref[Fig:Poisson:Morph]{\ref*{Fig:Poisson:Morph}(d)},~\hyperref[Fig:Poisson:Morph]{\ref*{Fig:Poisson:Morph}(e)}, and~\hyperref[Fig:Poisson:Morph]{\ref*{Fig:Poisson:Morph}(f)} show the evolution of the measured load $P$ as a function of the absolute displacement $\delta$. The former is the force measured at the end of the sample connected to the movable cross-head of the testing machine, the latter is the length of the sample during the unfolding process, as illustrated in Fig.~\ref{Fig:Poisson:Morph}(b).
The experimental (continuous lines) and numerical (dashed lines) curves were superimposed assuming as a common reference point at the sample rest configuration depicted in Fig.~\ref{Fig:Manufacturing}(a) right part. In such a configuration the load must be zero in both experiments and simulations. As expected, the load shows an inversion from compression (negative) to traction (positive) as it passes through the rest configuration during the unfolding process. The figure shows an almost prefect match between experimental results and numerical simulations for the folded state $n_m$=0. A small discrepancy is attained for the folded state $n_m$=1 when extreme folding is achieved. The issue is related to contact between the panels in the Miura mode due to their thickness, which occurs when high packing is achieved. As expected, the graphs show that the problem is emphasized in the case of $n_m$=2 and $n_m$=3 since the numbers of cells in Miura mode is increased. 

Poisson's ratio experiments were also performed on the standard Miura-ori and Eggbox patterns by using \textit{setup B} to validate the novel gripping systems in the case of standard origami patterns. Experiments on Miura-ori cannot be performed using \textit{setup A} due to its configuration shift from a 3D to a 2D and vice-versa during the folding/unfolding processes that leads to the slipping of the sample from the L-shaped profile. The experimental results (markers) are reported in Fig.~\ref{Fig:Poisson:Stand}(a) together with the theoretical prediction (continuous lines) and the simulations performed by MERLIN software (five-pointed star markers).
The colour of lines and markers indicate the investigated pattern type, Standard Eggbox red, and Standard Miura-ori blue.
Photos of the investigated patterns are reported in  Fig.~\ref{Fig:Poisson:Stand}(b), in which the areas highlighted in the middle of the samples indicate the tracking region exploited to determine the Poisson's ratio. The theoretical curves reported in the figure have been obtained from Eqns.~\ref{Eq:nu_Miura} (Miura-ori) and \ref{Eq:nu_Eggbox} (Eggbox).  An almost perfect agreement between theory, experiments, and numerical simulations is shown in the Fig.~\ref{Fig:Poisson:Stand}(a), thus confirming the efficiency of the novel setup. A video recording of the experiments is also provided as supporting material (Movie SM2).
The load $P$ as a function of the absolute displacement $\delta$ acquired during the experiments on the Miura-ori and the Eggbox patterns are reported in Figs.~\ref{Fig:Poisson:Stand}(c) and~\ref{Fig:Poisson:Stand}(d), respectively. The load $P$ is the force measured at the end of the sample connected to the movable cross-head of the testing machine, whereas the displacement $\delta$ represents the absolute length of the sample during the folding/unfolding process, as illustrated in Fig.~\ref{Fig:Poisson:Stand}(b). In each graph, the experiments (continuous lines) are compared with the MERLIN simulations (dashed lines). 
As previously discussed, experiments and simulations were superimposed assuming that  when the origami patterns are at rest configuration (depicted in Fig.~\ref{Fig:Manufacturing}(b) and~\ref{Fig:Manufacturing}(c) right part) the load must be zero. Also in this case, the load shows an inversion from compression (negative) to traction (positive) as it passes through the rest configuration during the unfolding process. The graphs show an almost prefect match between experimental results and numerical simulations.


\section{Concluding remarks}

We have conceived, designed, and realized a new setup to test deployable metamaterials without inducing boundary conditions frustration on the sample. The novel setup comprises an innovative gripping mechanism, called \textit{Saint-Venant fixture}, which eliminates the Saint-Venant end effect that usually arises during uniaxial testing experiments at the sample constrained regions. We validated this setup by performing Poisson's ratio experiments on different 2D origami tessellations, namely those composed of the Morph, the standard Miura-ori and the standard Eggbox patterns. We have demonstrated that the proposed gripping system allows a completely free sample deployment in both transverse and longitudinal directions, thus mimicking the theory's hypothesis. The coincidence between the experimental results,  the theory, and the MERLIN simulations has proved the high efficiency of the new gripping mechanism in the case of experiments involving patterns that exhibit extreme folding, from 2D to 3D (and vice-versa), such as is the case of the standard Miura-ori. The proposed experimental setup can also be extended for 3D metamaterials in general and adopted to study other tunable properties like programmable finite-strain Poisson's ratios~\citep{vasudevan2021origami}, elastic bandgaps~\citep{pratapa2018bloch}, thermal expansion~\citep{boatti2017origami}, which are strongly affected by boundary conditions.


\section*{Acknowledgement}
D.M.\,\,gratefully acknowledges the support by H2020- MSCA-ITN-2020-LIGHTEN-956547 grant and the Young Researchers GNFM 2020 grant. P.P.P.\,\,acknowledges the support from the Indian Institute of Technology Madras through the seed grant and the Science \& Engineering Research Board (SERB) of the Department of Science \& Technology, Government of India, through award SRG/2019/000999.

\appendix

\section{Poisson's ratio experiments by \textit{setup A}}\label{app:setup:A}

\begin{figure}[!ht]
	\centering
	\includegraphics[width=1\linewidth]{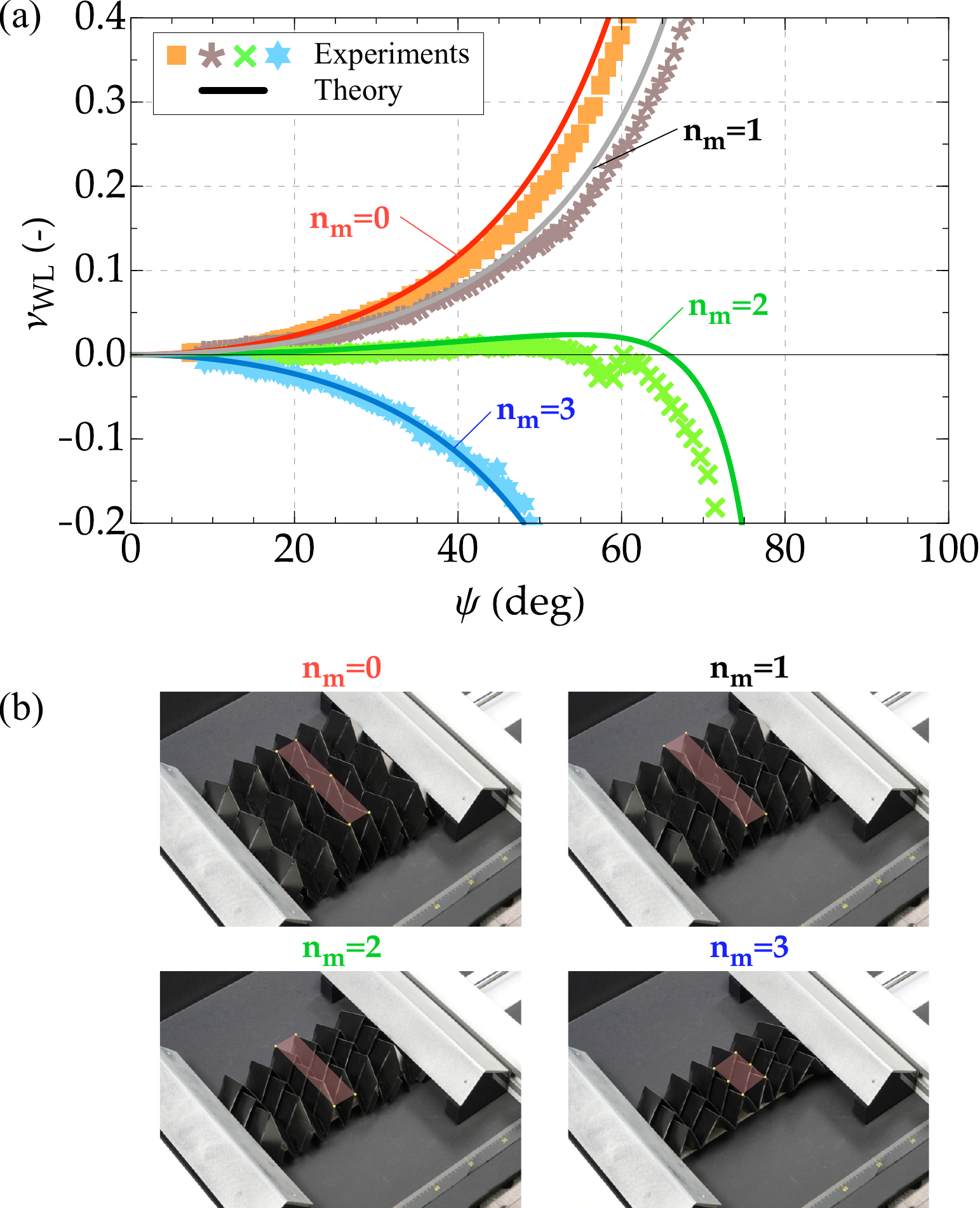}
	\caption{(a) Poisson's ratio as a function of the folding angle $\psi$ for the Morph pattern. In the graph, the experiments (markers) performed by using \textit{setup A} are compared with the theory (continuous lines). (b) Photos of the investigated folded states. The highlighted areas in the middle of the patterns indicate the tracking regions used to evaluate the Poisson's ratio during the experiments. }
	\label{Fig:Poisson:Morph:A}
\end{figure}

The first experimental campaign on the Morph pattern was carried out by using \textit{setup A}. Fig.~\ref{Fig:Poisson:Morph:A}(a) shows the Poisson's ratio evolution as a function of the angle $\psi$ for all the investigated hybrid states ($n_m=\{0,1,2,3\}$) depicted in Fig.~\ref{Fig:Poisson:Morph:A}(b). In the figure, the experimental results (markers) are compared with the theoretical prediction (continuous lines) provided by Eqn.~\ref{Eq:nuGlobal0}. The colour of lines and markers indicate the investigated folded state, $n_m$=0 red, $n_m$=1 dark gray, $n_m$=2 green, and $n_m$=3 blue.
Although the results obtained with the \textit{setup A}  are in good agreement with the theory up to an angle $\psi$ of about 45$^\circ$ they show a slight deviation from the predicted values while the samples undergo large unfolding. 
The cross-check of movies and experimental data shown in  Fig.~\ref{Fig:Poisson:Morph:A}(a) indicates the gripping system adopted in \textit{setup A} as the main reason for the discrepancy between theory and experiments. The L-shaped profiles used to connect the sample to the testing machine limit the transverse deployment of the metamaterial, inducing stress concentration on the sample in the constrained region. This leads to a non-uniform transverse deformation to the sample giving rise to a dog-bone shape. Moreover, such a gripping system reduces also the dimension of the tested patterns from 8$\times$3 to 6$\times$3, holding the rows on the two ends (one on each of the shorter sides) beneath the L-shaped profile, reducing the region in the middle of the sample without boundary effects.  For these reasons, we have designed, realized and tested a completely new and innovative setup, the \textit{setup B}, based on a linear slide rail system that leaves completely free the deployment of the origami metamaterials in the transverse direction.  

\section{Numerical simulations using the MERLIN software}\label{app:MERLIN}
\noindent The numerical simulations are performed using the MERLIN software (version 2) \citep{liu2017nonlinear,liu2018merlin2}. The software implements the bar-and-hinge model as a reduced order model of origami structures. In particular, the N5B8 model \citep{filipov2017bar} discretizes each quadrilateral panel into four triangles and represents the origami behavior by bars and rotational springs. The N5B8 model captures three essential deformation modes of origami structures: folding, panel bending, and in-plane stretching. The deformation of the origami structure is obtained by solving the nonlinear equilibrium equation, i.e., finding the stationary states of the system energy \citep{liu2017nonlinear}. Such problem needs to be solved by an incremental-iterative numerical algorithm, for which we adopt the Modified Generalized Displacement Control Method \citep{liu2017nonlinear,liu2018merlin2}. In this work, we use the actual Young's modulus ($E$=1.4 GPa), Poisson's ratio ($\nu$=0.36), and thickness ($t$=1.0 mm) of the Polypropylene sheet. The constitutive model of the folding hinges are assumed to follow the enhanced linear model as proposed in reference \citep{liu2017nonlinear}:
\begin{equation} 
  \frac{M(\theta)}{L_F} = \left\{
 \begin{array}{lr}
 k_0 (\theta_s-\theta_0) + \left( 2 k_0 \theta_s/\pi \right) \tan \left( \dfrac{\pi (\theta-\theta_s)}{2 \theta_s} \right),  0 < \theta < \theta_s; \\[3mm]
k_0 (\theta - \theta_0),   \hspace{31mm}   \theta_s \leqslant \theta \leqslant 2\pi - \theta_s; \\[3mm]
 k_0 (2\pi-\theta_s-\theta_0) + \\ 
  \left( 2 k_0 \theta_s/\pi \right) \tan \left( \dfrac{\pi (\theta+\theta_s-2\pi)}{2\theta_s} \right),  \hspace{2mm} 2\pi - \theta_s < \theta < 2 \pi.
\end{array}
\right.
\label{eqn:simple_one} 
\end{equation}

The linear stiffness $k_0$ of the folding hinges are determined by the length scale factor as explained in reference \cite{filipov2015origami,liu2018merlin2}, a phenomenological parameter that sets the relative stiffness of folding hinges to the bending stiffness of panels. In our study, for each sample, the stiffening angle ($\theta_s$) and the ratio of length scale factor to the length of a folding hinge ($L^*/L_F$) are tuned for best match with the experimental load-displacement curves. For the morph pattern, we assign $L^*/L_F=55$ and $\theta_s = 82^\circ$. For the Miura pattern, we assign $L^*/L_F=25$ and $\theta_s = 60^\circ$. For the eggbox pattern, we assign $L^*/L_F=18$ and $\theta_s = 65^\circ$.

\bibliography{references}

\end{document}